\documentclass[aps,floats,nofootinbib]{revtex4}
\usepackage{color}
\usepackage{amsmath,amssymb}
\usepackage{epsfig}

\def\lsim{\mathrel{\raise.3ex\hbox{$<$\kern-.75em\lower1ex\hbox{$\sim$}}}}
\def\gsim{\mathrel{\raise.3ex\hbox{$>$\kern-.75em\lower1ex\hbox{$\sim$}}}}

\newcommand{\nc}{\newcommand}
\nc{\beq}{\begin{equation}}  \nc{\eeq}{\end{equation}}
\nc{\bea}{\begin{eqnarray}}  \nc{\eea}{\end{eqnarray}}
\nc{\baa}{\begin{array}}     \nc{\eaa}{\end{array}}
\nc{\bit}{\begin{itemize}}   \nc{\eit}{\end{itemize}}
\nc{\ben}{\begin{enumerate}} \nc{\een}{\end{enumerate}}
\nc{\bce}{\begin{center}}    \nc{\ece}{\end{center}}
\nc{\bpm}{\begin{pmatrix}}   \nc{\epm}{\end{pmatrix}}
\nc{\bvt}{\begin{verbatim}}  \nc{\evt}{\end{verbatim}}

\newcommand{\be}{\begin{eqnarray}}
\newcommand{\ee}{\end{eqnarray}}
\newcommand{\nn}{\nonumber\\}
\def \siggaga     {$\sigma^{\gamma\gamma}$ }

\def\a               {\alpha}
\def\b               {\beta}
\def\l               {\lambda}

\nc{\bew}{\beta_W}
\nc{\dtz}{d^2_{0,0}}
\nc{\dzz}{d^0_{0,0}}
\nc{\dto}{d^2_{1,0}}

\def\bea{\begin{eqnarray}}
\def\eea{\end{eqnarray}}

\def\beq{\begin{equation}}
\def\eeq{\end{equation}}

\def\bce{\begin{center}}
\def\ece{\end{center}}

\def\hbar{\overline h}

\begin{document}
\title{$H\to\gamma\gamma$ in the Complex Two Higgs Doublet Model}
\author{Abdesslam Arhrib}
\affiliation{D\'epartement de Math\'ematique, Facult\'e des
Sciences et Techniques, Universit\'e Abdelmalek Essa\^adi,
B.~416, Tangier, Morocco}

\author{Rachid Benbrik}
\affiliation{Instituto de F\'isica de Cantabria (CSIC-UC), Santander, Spain.}
\affiliation{Facult\'e Polydisciplinaire de Safi, Sidi Bouzid B.P 4162, 46000 Safi,
Morocco.}
\affiliation{LPHEA, FSSM, Cadi Ayyad University, B.P. 2390, Marrakesh,
  Morocco.}

\author{Chuan-Hung Chen}

\affiliation{ 
Department of Physics, National Cheng-Kung University, Tainan 701, Taiwan.}
\affiliation{National Center for Theoretical Sciences, Hsinchu 300, Taiwan.}
\begin{abstract}
We explore the parameter regions in the complex Two Higgs Doublet Model
(c2HDM) with explicit CP violation  that can  describe current LHC hints 
 for one or more Higgs boson signals at 125 GeV.
Such a simple extension of the Standard model has three
  neutral Higgs bosons and a pair charged Higgs and  leads to rich
  CP-violating sources including the CP-even CP-odd mixing of the neutral 
Higgs bosons.
Within this model we present the production of light
  Higgs boson at the LHC followed by its decay into two photons. Our numerical
  study takes into account theoretical and experimental 
constraints on the Higgs potential like
positivity, unitarity and perturbativity as well as
  $\rho$-parameter, $b\to s\gamma$ and $R_b$. 
These requirement together with the
minimum conditions could explain the di-photon excess observed at the LHC. 
We also discuss the effects of the CP violating phases on the CP violation 
observable which is the difference between left-circular
and right-circular polarization.
\end{abstract}

\maketitle


\section{Introduction}
\label{intro}
LHC experiments at 7 TeV after analyzing  their 5/fb
dataset restrict the Standard Model (SM) light Higgs boson 
mass to be in the following range 115.5 GeV $< m_h <$ 131 GeV for ATLAS
\cite{atlas} and $m_h <$ 127 GeV for CMS \cite{cms} at 95\% 
confidence level (CL). 
Both  ATLAS and CMS have reported some excess at low mass
Higgs Boson with low  statistical significance in the $WW^*$,
$ZZ^*$ and di-photon channels. Moreover, from the di-photon channel, 
both detectors have excluded a SM Higgs in  the narrow mass range of
$114$--$115$~GeV for ATLAS and  of
$127$--$131$ GeV for CMS at the $95 \%$C.L.
Moreover, in the mass range $115< m_h < 135$ GeV, Tevatron also observes 
the excess events in $h\to b \bar b$ decay \cite{hbb}.

LHC is now running at 8 TeV, the hope is to reach more than 5/fb 
luminosity taken during 7 TeV run. This would allow the LHC experiments 
to take more and more data and then could elucidate the existing hint
 at 125 GeV.\\
The effective cross-section of di-photon ($\gamma \gamma$) mode can
 be estimated by inclusive process $\sigma^{\gamma\gamma} = 
\sigma(p p \to H ) \times Br(H \to \gamma \gamma)$. 
This (\siggaga) could provide possibly
the best mode to search for light Higgs Boson in mass range 110-140 GeV. 
ATLAS \cite{atlas_diphoton} reported 95\% CL exclusion limit of 
$\sigma^{\gamma\gamma}/\sigma^{\gamma\gamma}_{SM} \sim 1.6 - 1.8$ in
mass range 110-130 GeV. On the other hand,
CMS  \cite{cms_diphoton}  reported the exclusion limit of  
$\sigma^{\gamma\gamma}/\sigma^{\gamma\gamma}_{SM} \sim 1.5 - 2$ in
mass range 110-140 GeV. 

If we take the result of 
$\sigma^{\gamma\gamma}/\sigma^{\gamma\gamma}_{SM} > 1$ 
seriously, it seems that LHC does not only observe a scalar 
particle, but also indicate the necessity to extend the SM. 
One of the simplest extension of the SM is to add an extra
  Higgs doublet which is motivated from spontaneous CP-violation. 
It has been shown  that if one Higgs doublet is needed for 
  the mass generation, then the extra
  Higgs doublet is necessary for the spontaneous CP-violation yielding an
  extremely rich array of possible phenomenological consequences. To mention a
few, the new sources of CP-violation can generate observable CP-asymmetries at
the LHC as well as at future linear collider. A large CP-violating phases
could have an effect on many CP-conserving observables such the production
rate and the decay ratios of Higgs bosons.

Recently, several studies has been done for higgs decay into 2 photons 
in the framework of extended Higgs sector \cite{2HDMp}.  
In the present work, we assume that the Higgs is produced dominantly by
  gluon fusion, the signal strength is defined by the ratio
\begin{eqnarray}
R_{XX} (h_1) = [\sigma(pp\to h_1)\times Br(h_1\to
XX)]/[\sigma(pp\to h)\times Br(h\to XX)]_{SM}
\label{ratioxx}
\end{eqnarray}
where $X$ can be either photons,$W$ or $b$. In this ratio, we have used the narrow width approximation.\\
In our calculations, we take into account positivity, unitarity and 
perturbativity of the Higgs potential as well as $\rho$-parameter, 
$b\to s\gamma$ and $R_b$. Since the complex Yukawa couplings are 
chosen in the calculations, in addition to the CKM-matrix elements, 
 there is a new source for CP-violation. Using those new 
CP violating sources,  we obtain $R_{\gamma\gamma} \sim 3$ and a rate 
difference between left-circular and right-circular polarization 
$A_{+-}$ of the order of 10\%.

The paper is organized as follows: In section.II we shortly review the
  general 2HDM, fix the parameterization for the scalar potential. 
In the third  section  we investigate how each
phase can affect $R_{\gamma\gamma}$ , we study the correlations 
between the various $R_{\gamma\gamma}$, $R_{WW}$ and $R_{BB}$, we then
discuss the phenomenology implications of such phases on $A_{+-}$. 
Finally, we devote section IV to our summary and conclusions.


\vspace*{-.2in}
\section{General c2HDM}
\label{res}
\vspace*{-.1in}
The general 2HDM is obtained via extending the SM Higgs sector,
consisting of one complex Y= +1,
$\rm{SU (2)_L}$ doublet scalar field $\Phi_1$, with an additional complex
Y = +1, $\rm{SU (2)_L}$ doublet scalar field $\Phi_2$.
Using $\Phi_{1,2}$, one can build the most
general renormalizable $\rm{SU(2)_L \times U(1)_Y}$ gauge
invariant Higgs potential \cite{THDM_CPC1,THDM_CPC2,THDM_CPC22,THDM_CPC23,tdlee,tdlee2,tdlee3,tdlee4,Arhrib:2010ju}:
\begin{eqnarray}
\hspace*{-1cm}
 V_{\rm{Higgs}}(\Phi_1,\Phi_2) = \frac{\l_1}{2}(\Phi_1^\dagger\Phi_1)^2 +
\frac{\l_2}{2}(\Phi_2^\dagger\Phi_2)^2 +
\l_3(\Phi_1^\dagger\Phi_1)(\Phi_2^\dagger\Phi_2) +
\l_4(\Phi_1^\dagger\Phi_2)(\Phi_2^\dagger\Phi_1) \nn  +~
\frac12\left[\l_5(\Phi_1^\dagger\Phi_2)^2 +\rm{h.c.}\right]
+~\left\{\left[\l_6(\Phi_1^\dagger\Phi_1)+\l_7(\Phi_2^\dagger\Phi_)\right]
(\Phi_1^\dagger\Phi_2)+\rm{h.c.}\right\} \nn
-~\frac{1}{2}\left\{m_{11}^2 \Phi_1^\dagger \Phi_1+ \left[m_{12}^2
\Phi_1^\dagger \Phi_2 + \rm{h.c.}\right]
 + m_{22}^2\Phi_2^\dagger
\Phi_2 \right\}\,.\label{CTHDMpot}
\end{eqnarray}
By hermiticity of eq. (\ref{CTHDMpot}),
$\l_{1,2,3,4}$, as well as $m_{11}$ and $m_{22}$ are real-valued;
while the dimensionless parameters $\l_{5}$, $\l_6$, $\l_7$ and $m_{12}^2$ are in general complex.

\subsubsection{Mass eigenstates}
After the $\rm{SU(2)_L \times U(1)_Y}$ gauge symmetry is broken down
to $\rm {U (1)_{em}}$ via the Higgs mechanism, one can choose a basis
where the vacuum expectation values (VEVs) of the two Higgs doublets, 
$v_1$ and $v_2$ are non-zero, real and positive, and fix the following
parameterization \cite{Maria}:
\be
 \Phi_1= {\small \left(
  \begin{array}{c}
   \varphi_1^+\\ (v_1+\eta_1+i\chi_1)/\sqrt{2}
  \end{array}\right)}\,,
\quad \Phi_2= {\small \left( \begin{array}{c}
  \varphi_2^+\\ (v_2+\eta_2+i\chi_2)/\sqrt{2}
  \end{array}\right)}\,.\label{Hdoublets}
\ee
Here $\eta_{1,2}$ and $\chi_{1,2}$ are neutral scalar fields
and $\varphi_{1,2}^\pm $ are charged scalar fields.
The physical Higgs eigenstates are obtained as follows.

The charged Higgs fields $H^\pm$ and the
charged would-be Goldstone boson fields $G^\pm$ are a mixture of the
charged components of the Higgs doublets  (\ref{Hdoublets}), 
$\varphi^\pm_{1,2}$:
\begin{eqnarray}
H^\pm &=& -\sin\beta\varphi^\pm_1+\cos\beta\varphi^\pm_2\,,
\nn
G^\pm &=& \cos \beta \varphi^\pm_1+\sin\beta\varphi^\pm_2\,,
\end{eqnarray}
where the mixing angle $\beta$ is defined through the ratio of the
VEVs of the two Higgs doublets $\Phi_2$ and $\Phi_1$, $\tan\beta=v_2/v_1$. 
$G^\pm$ give masses to the $W^\pm$ bosons.

Obtaining the neutral physical Higgs states is a few steps procedure.
First, one rotates the
imaginary parts of the neutral components of eq. (\ref{Hdoublets}): 
$(\chi_1,\chi_2)$ into the basis $(G^0,\eta_3)$:\footnote{Note that in 
the case of $m_{12}^2=\l_{6}=\l_7=0$ and all other parameters of 
eq. (\ref{CTHDMpot}) are real, the physical Higgs sector of the 
2HDM is analogous to the one of the tree-level MSSM. In this case 
the scalar field $\eta_3$ is equivalent to the  MSSM neutral CP-odd 
Higgs boson $A^0$.}
\begin{eqnarray}
G^0 &=& \cos \beta \chi_1+\sin\beta\chi_2\,, \nn
\eta_3 &=& -\sin\beta\chi_1+\cos\beta\chi_2\,,
\label{CPoddHiggs}
\end{eqnarray}
where $G^0$ is the would-be Goldstone boson which
gives a mass to the $Z$ gauge boson.
After elimination of the Goldstone mode, the remaining
neutral CP-odd component $\eta_3$ mixes with the neutral CP-even
components $\eta_{1,2}$. The relevant squared mass
matrix ${\cal M}^2_{ij}=\partial^2 V_{\rm{Higgs}}/(\partial
\eta_i\partial \eta_j )$, $i,j=1,2,3$,
has to be rotated from the so called "weak basis" 
$(\eta_{1},\eta_{2},\eta_{3})$ to the diagonal basis
$(H_1^0, H_2^0, H_3^0)$ by an orthogonal $3\times 3$ 
matrix ${\cal{R}}$ as follows:
\be
{\cal R} {\cal M}^2 {\cal R}^T={\cal
M}^2_{\rm{diag}}={\rm diag}({\rm M}_{H_1^0}^2, {\rm M}_{H_2^0}^2, 
{\rm M}_{H_3^0}^2)\,, \label{mixmat}
\ee
with
\be
{\small \left(
  \begin{array}{c}
   H_1^0\\ H_2^0 \\H_3^0
  \end{array}\right)}=
{\cal R} {\small \left( \begin{array}{c}
  \eta_1\\ \eta_2 \\ \eta_3
  \end{array}\right)}\,,
\ee
where we have defined the Higgs fields $H_i^0$ such that 
their masses satisfy the
inequalities:
\be
\rm{M_{H_1^0}}\leq\rm{M_{H_2^0}}\leq\rm{M_{H_3^0}}\,.
\ee
Note that the mass eigenstates $H_i^0$ have a mixed CP structure.

Following \cite{ElKaffas},
we parameterize the orthogonal $3\times 3$ matrix
${\cal{R}}$ by three rotation angles ~$\alpha_i,~
i=1,2,3$:
\begin{eqnarray}
\cal{R} &=& {\small \left(
  \begin{array}{ccc}
   1         &    0         &    0
\\ 0 &  \cos\alpha_3 & \sin\alpha_3 \\ 0 & -\sin\alpha_3 &
\cos\alpha_3
  \end{array}\right)}
 {\small \left(
  \begin{array}{ccc}
\cos\alpha_2 & 0 & \sin\alpha_2 \\ 0         &       1         & 0
\\ -\sin\alpha_2 & 0 & \cos\alpha_2
  \end{array}\right)}
   {\small \left(
  \begin{array}{ccc}
\cos\alpha_1 & \sin\alpha_1 & 0 \\ -\sin\alpha_1 & \cos\alpha_1 &
0 \\ 0         &       0         & 1
  \end{array}\right)}\nn
 &=& {\small \left(
  \begin{array}{ccc}
   c_1\,c_2 & s_1\,c_2 & s_2 \\ - (c_1\,s_2\,s_3 + s_1\,c_3) & c_1\,c_3 -
s_1\,s_2\,s_3 & c_2\,s_3 \\ - c_1\,s_2\,c_3 + s_1\,s_3 & -
(c_1\,s_3 + s_1\,s_2\,c_3) & c_2\,c_3
  \end{array}\right)}\,,
  \label{matRo}
\end{eqnarray}
with $s_i=\sin\alpha_i$ and $c_i=\cos\alpha_i$, which we vary in our numerical analysis in the following ranges:
\be
-\frac{\pi}{2} < \a_1\leq \frac{\pi}{2}~; \quad -\frac{\pi}{2} <
\a_2\leq \frac{\pi}{2}~; \quad 0 \leq \a_3\leq \frac{\pi}{2}~.
\ee
Note that in the limit of no CP violation $\alpha_{2,3} \to 0
  (R_{13}, R_{23} \to 0)$
  in this case the neutral Higgs sector is parameterized by the familiar mixing
  angle $\alpha_1$ of the CP-even sector. For consistency, this requires
  $Im(\lambda_5)$ and $Im(m^2_{12})$ to be zero. 
For the next section we will study the general 2HDM where 
$\lambda_6=\lambda_7 =0$.

\subsection{$Z_2$ symmetry and input parameter set}
\label{sec:Z2param}

In the most general 2HDM, some types of Yukawa interactions
can introduce flavor changing neutral currents
(FCNC) already at tree level. It is well known that the latter effects are small in nature.
This problem has been solved by imposing a discrete $Z_2$ symmetry on the Lagrangian. It forbids
 $\Phi_1\leftrightarrow\Phi_2$ transitions and in its exact form it also leads
 to conservation of CP \cite{Glashow}. In order to allow some effects of CPV
it is necessary to violate the $Z_2$ symmetry. Basically, there are two ways of $Z_2$ symmetry violation -- "soft"
 and "hard". A softly broken $Z_2$ symmetry suppresses FCNC at tree level, but still allows CPV.

In this paper we will work in a model of a softly broken $Z_2$ 
symmetry of the 2HDM Lagrangian.
This forbids the quartic terms proportional to $\l_6$ and $\l_7$ in eq. (\ref{CTHDMpot}), but the quadratic term with $m^2_{12}$ is still allowed
 \cite{Khater}:
\begin{eqnarray}
\hspace*{-2cm}
 V_{\rm{Higgs}}^{\rm soft}(\Phi_1,\Phi_2) = \frac{\l_1}{2}(\Phi_1^\dagger\Phi_1)^2 +
\frac{\l_2}{2}(\Phi_2^\dagger\Phi_2)^2 +
\l_3(\Phi_1^\dagger\Phi_1)(\Phi_2^\dagger\Phi_2) +
\l_4(\Phi_1^\dagger\Phi_2)(\Phi_2^\dagger\Phi_1) \nonumber \\  +~
\frac12\left[\l_5(\Phi_1^\dagger\Phi_2)^2 +\rm{h.c.}\right]
-~\frac{1}{2}\left\{m_{11}^2 \Phi_1^\dagger \Phi_1+ \left[m_{12}^2
\Phi_1^\dagger \Phi_2 + \rm{h.c.}\right]
 + m_{22}^2\Phi_2^\dagger
\Phi_2 \right\}\,.\label{THDMpot}
\end{eqnarray}

The Higgs potential ~(\ref{THDMpot}) has 12 real parameters: 2 real masses: $m^2_{11,\,22}$, 2 VEVs: $v_{1,\,2}$, four real quartic couplings: $\l_{1,\,2,\,3,\,4}$ and two complex parameters: $\l_5$ and $m_{12}^2$. The conditions for having an extremum of eq. (\ref{THDMpot}) reduce the number of parameters:
$m^2_{11\,,22}$ are eliminated by the minimization conditions, and the combination $v_1^2+v_2^2$ is fixed at the electroweak scale $v=(\sqrt 2 G_F)^{-1/2}$ = 246 GeV. Moreover, in this case the minimization conditions also relates ${\rm Im}\, (m_{12}^2)$ and ${\rm Im}\, (\l_5)$:
 \be
{\rm Im}\, (m_{12}^2) = v_1\,v_2\, {\rm Im}\,(\l_5)\,. \label{oneCP}
 \ee
Thus, our Higgs potential  (\ref{THDMpot}) is a function of 8 real independent parameters:
 \be
\big\{\l_{1,2,3,4},~~ {\rm Re}\, (\l_5),~~ {\rm Re}\,
(m_{12}^2),~~\tan \beta,~~ {\rm Im}\, (m_{12}^2)\big\}. \label{set1}
\ee
It contains minimal CPV generated by $m_{12}^2 \ne 0$ and complex.
In our further analysis we will use the following parameter set equivalent to eq. (\ref{set1}):
\be
\bigg\{M_{H_1^0},~~ M_{H_2^0},~~ M_{H^+},~~ \alpha_1,~~
\alpha_2,~~ \alpha_3,~~\tan \beta,~~ {\rm Re}\, (m_{12})
\bigg\}\,.\label{paramm}
\ee
Note that the mass of the heaviest neutral Higgs boson $H_3^0$ 
is not an independent parameter.
In the considered CP violating case, the matrix elements $({\cal M}^2)_{13}$
and $({\cal M}^2)_{23}$ of the squared mass
matrix  (\ref{mixmat}) are non-zero and correlated \cite{theor.constr.}:
\be
({\cal M}^2)_{13}&=&\tan \b \,({\cal M}^2)_{23}\,.
\label{sumrule0}
\ee
Writing this relation in terms of the physical masses
${\rm M}_{H_1^0}$, ${\rm M}_{H_2^0}$, ${\rm M}_{H_3^0}$ one obtains \cite{theor.constr.}:
\be
M_{H_3^0}^2=\frac{M_{H_1^0}^2 R_{13}(R_{12}\tan
\beta-R_{11})+M_{H_2^0}^2R_{23}(R_{22}\tan
\beta-R_{21})}{R_{33}(R_{31}-R_{32}\tan \beta)}\,,\label{M3}
\ee
where $R_{ij}\,, i,j=1,2,3,$ are the elements of the rotation 
matrix (\ref{matRo}).

The expressions for the parameters $\l_{1,2,3,4},~ {\rm Re} \l_5, 
{\rm Im} \l_5$ of the scalar potential  (\ref{THDMpot}) 
as a functions of the physical masses and mixing angles are 
given in \cite{ElKaffas}. Note that several CP conserving limit exists
depending on which Higgs $H_i$ is pure CP-odd we can set different limit.
If we want $H_3$ to be the CP-odd then we take 
  $\alpha_2=\alpha_3=0$, and $\alpha_1$ arbitrary.


\begin{figure}[!ht]
\begin{tabular}{ccc}\hspace*{-1cm}
\resizebox{69mm}{!}{\includegraphics{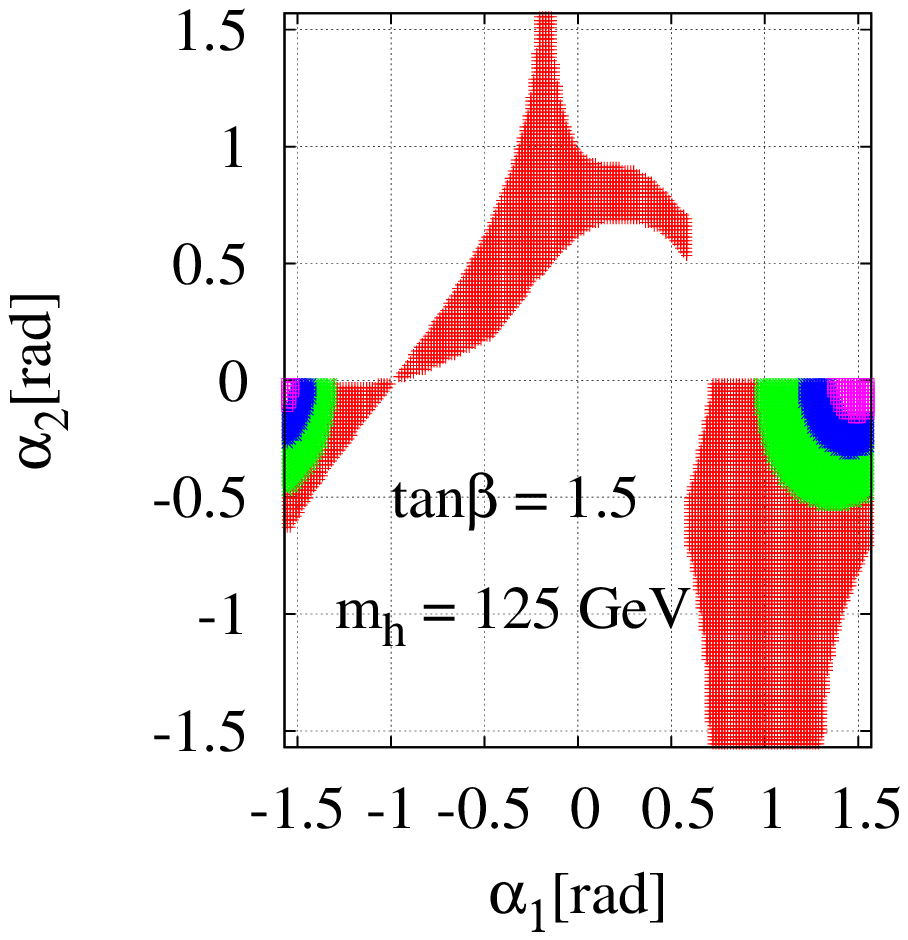}} & \hspace{-2cm}
\resizebox{69mm}{!}{\includegraphics{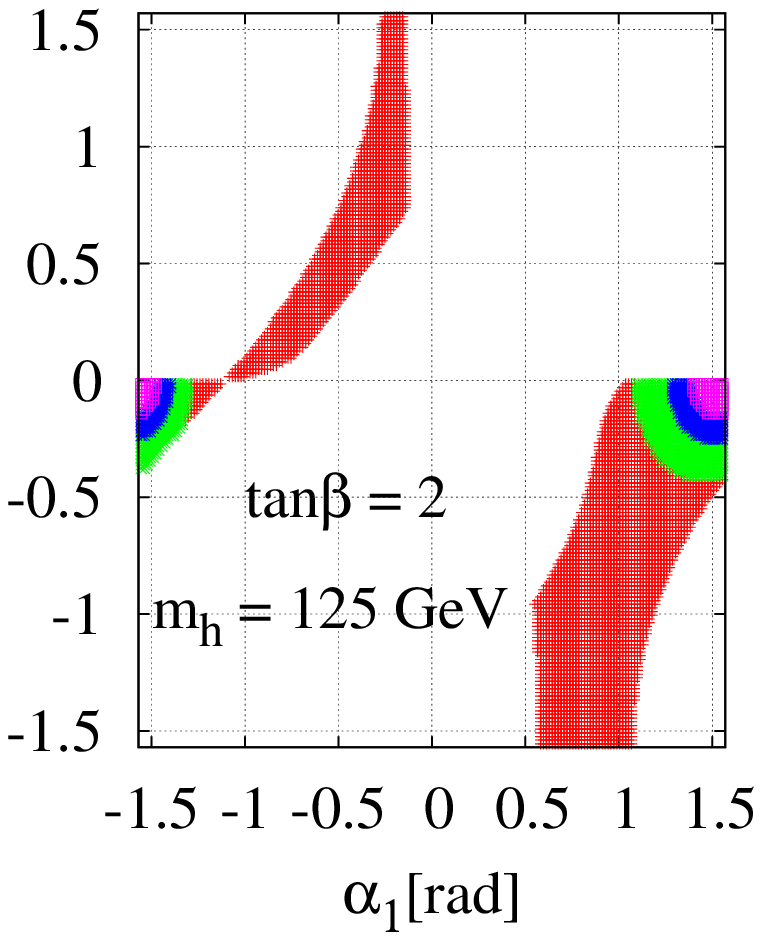}} & \hspace{-2cm}
\resizebox{69mm}{!}{\includegraphics{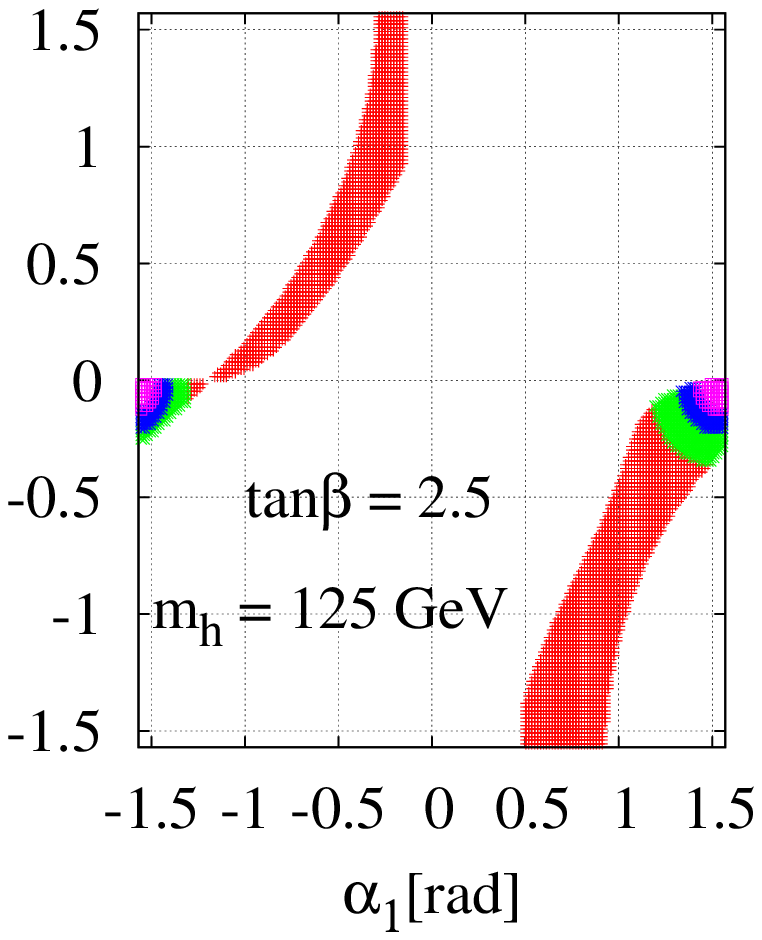}}
\end{tabular}
\caption{Allowed parameter space in the $\alpha_1 - \alpha_2$ plane, for
  $\tan\beta = 1.5 $ (left), $\tan\beta = 2$ (middle) and $\tan\beta = 2.5$
  (right). The coding color correspond to: $R_{\gamma\gamma} \le 1$
  (red),$1\le R_{\gamma\gamma} \le 2$ (green),$2\le R_{\gamma\gamma} \le 3$
  (blue) and $3\le R_{\gamma\gamma}$ (mangeta). The other parameters are taken as follow: $m_{h_1} = 125$ GeV,
  $m_{h_2} = 220$ GeV,  $m_{H^\pm} = 350$ GeV, $Re(m^2_{12}) = 200$ GeV,
  $\alpha_3 = \frac{\pi}{3}$. }
\label{figu:1}
\end{figure}

\begin{figure}[!ht]
\begin{tabular}{cc}\hspace*{-1cm}
\resizebox{69mm}{!}{\includegraphics{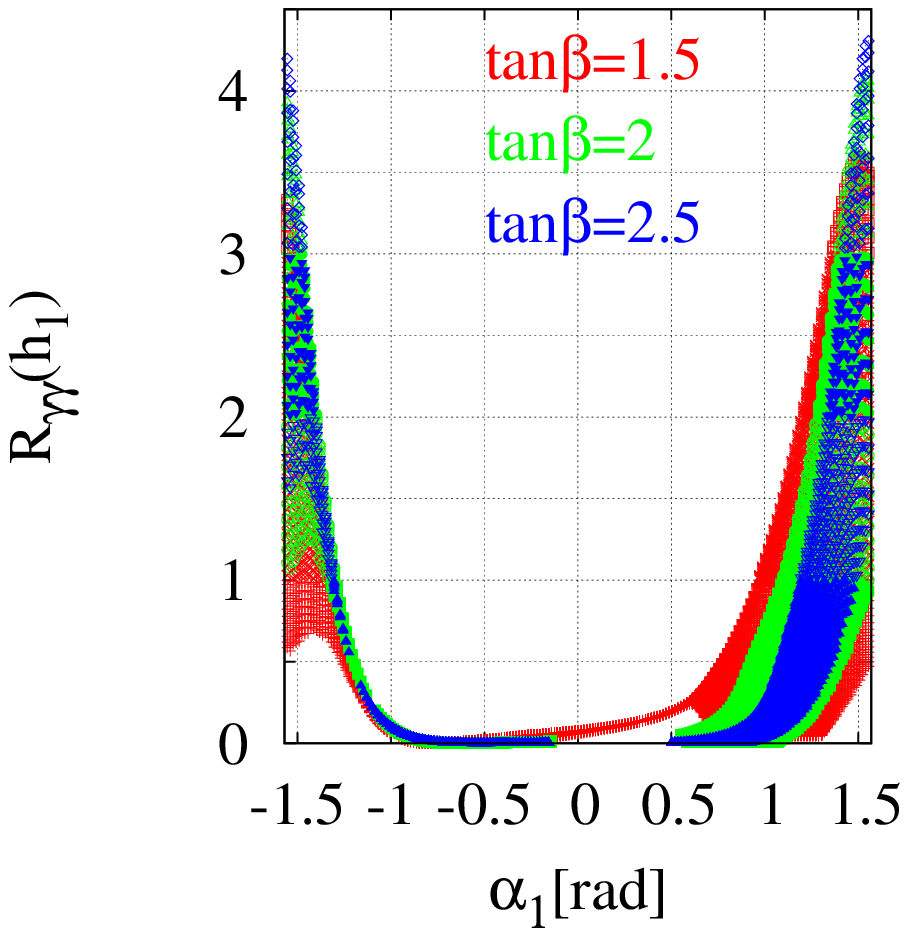}} & \hspace{-2cm}
\resizebox{69mm}{!}{\includegraphics{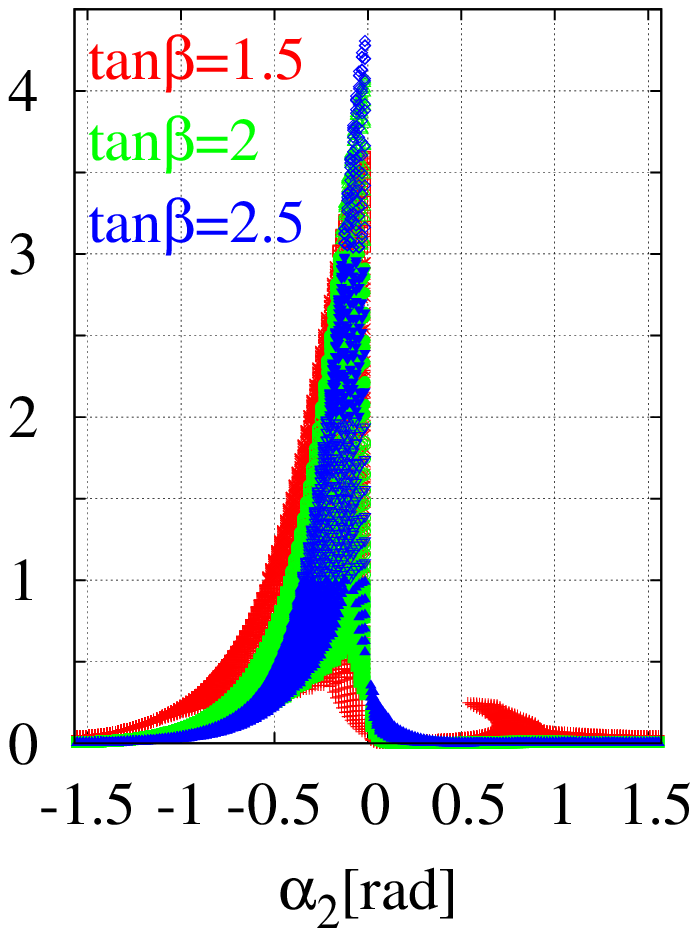}} 
\end{tabular}
\caption{$R_{\gamma\gamma}(h_1)$ as a function of  $\alpha_1$ (left) and 
$\alpha_2$ (right), for  different values of $\tan\beta = 1.5, 2$ and 
2.5. The other parameters are taken as follow: $m_{h_1} = 125$ GeV,
 $m_{h_2} = 220$ GeV,  $m_{H^\pm} = 350$ GeV, $Re(m^2_{12}) = 200$ GeV,
 $\alpha_3 = \frac{\pi}{3}$.}
\label{figu:2}
\end{figure}

\begin{figure}[!ht]
\begin{tabular}{c}\hspace*{-1cm}
\resizebox{95mm}{!}{\includegraphics{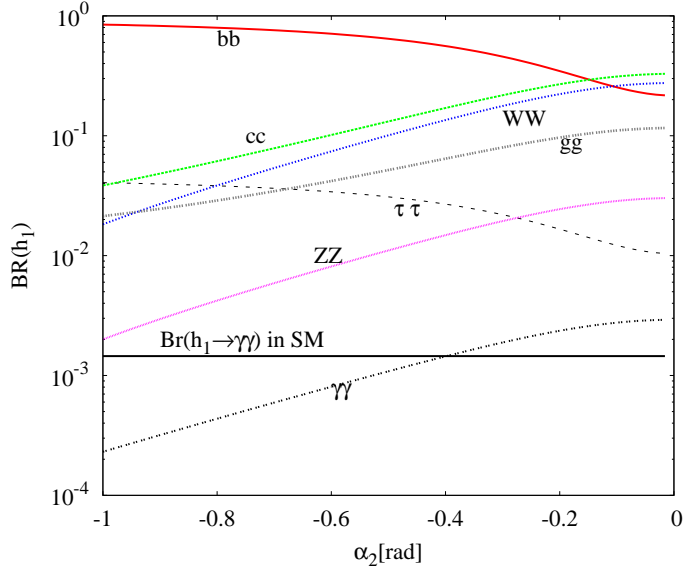}} 
\end{tabular}
\caption{Branching ratios of $h_1$ in the c2HDM as a function $\alpha_2$ with
  $\alpha_1 = 1.4 $ rad, $\alpha_3 = 1.04$ rad, $\tan\beta = 1.5$ and $m_{h_1}
  = 125$ GeV. The other parameters are as in Fig.1 }
\label{figu:3}
 \end{figure}

\begin{figure}[!ht]
\begin{tabular}{cc}\hspace*{-1cm}
\resizebox{69mm}{!}{\includegraphics{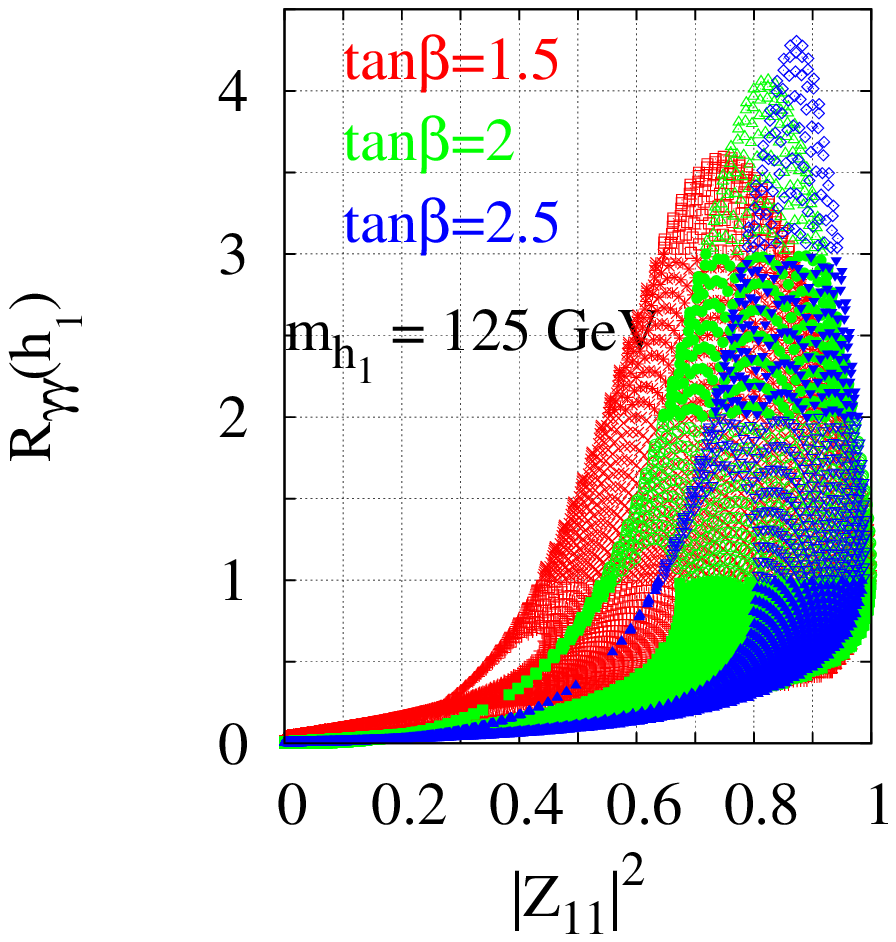}} & \hspace{-1cm}
\resizebox{69mm}{!}{\includegraphics{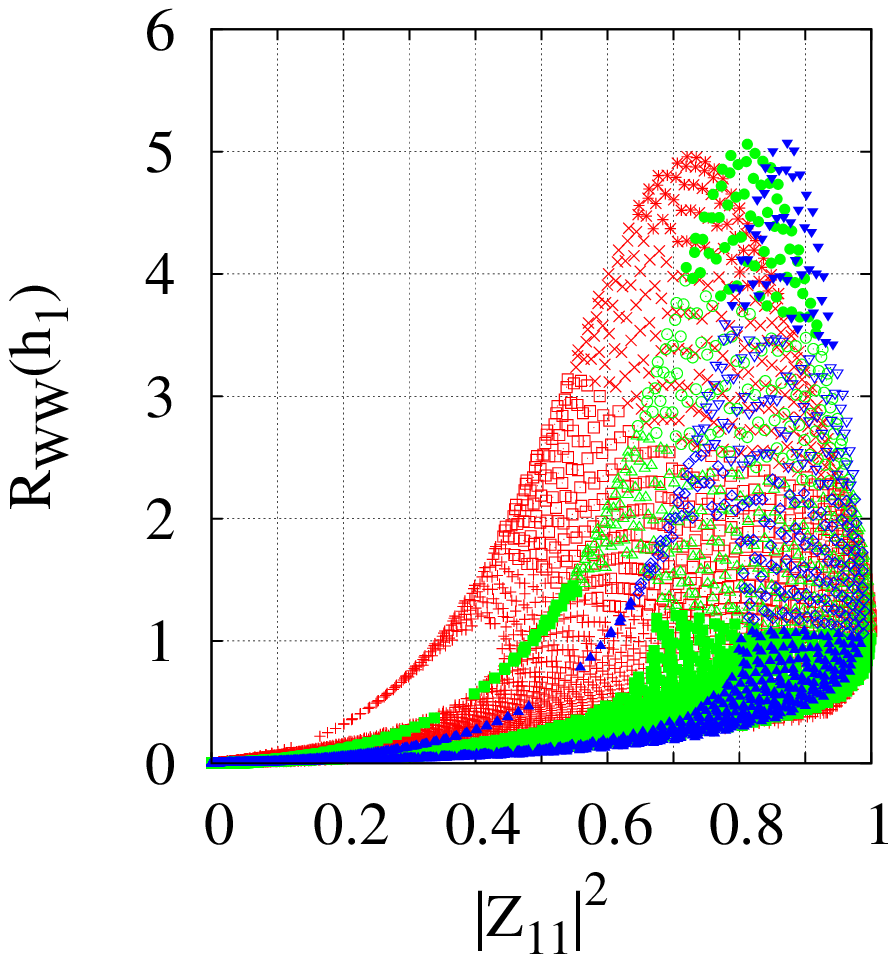}} 
\end{tabular}
\caption{$R_{\gamma\gamma}(h_1)$ as a function of reduced coupling 
$ Z_{11} = g_{h_1 WW}/g_{h_{sm}WW}$ in
  the c2HDM. The other parameters are the same as in Figure.1}
\label{figu:4}
 \end{figure}

\begin{figure}[!ht]
\begin{tabular}{cc}\hspace*{-1cm}
\resizebox{69mm}{!}{\includegraphics{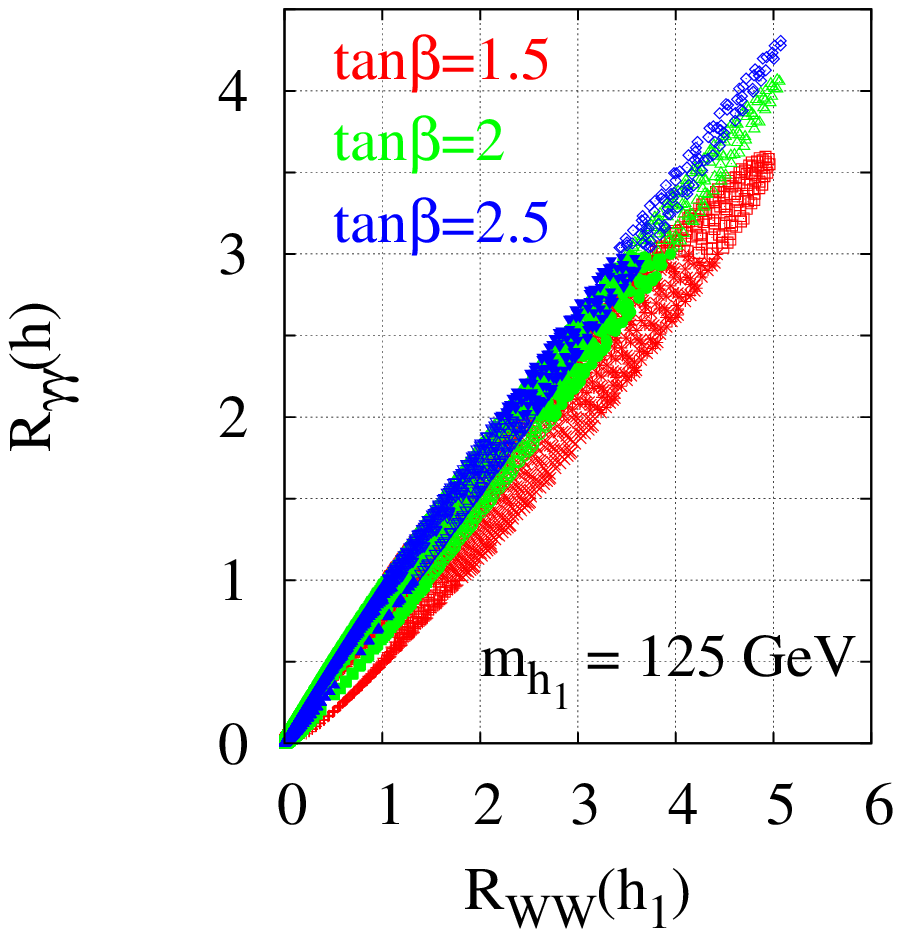}} & \hspace{-1cm}
\resizebox{69mm}{!}{\includegraphics{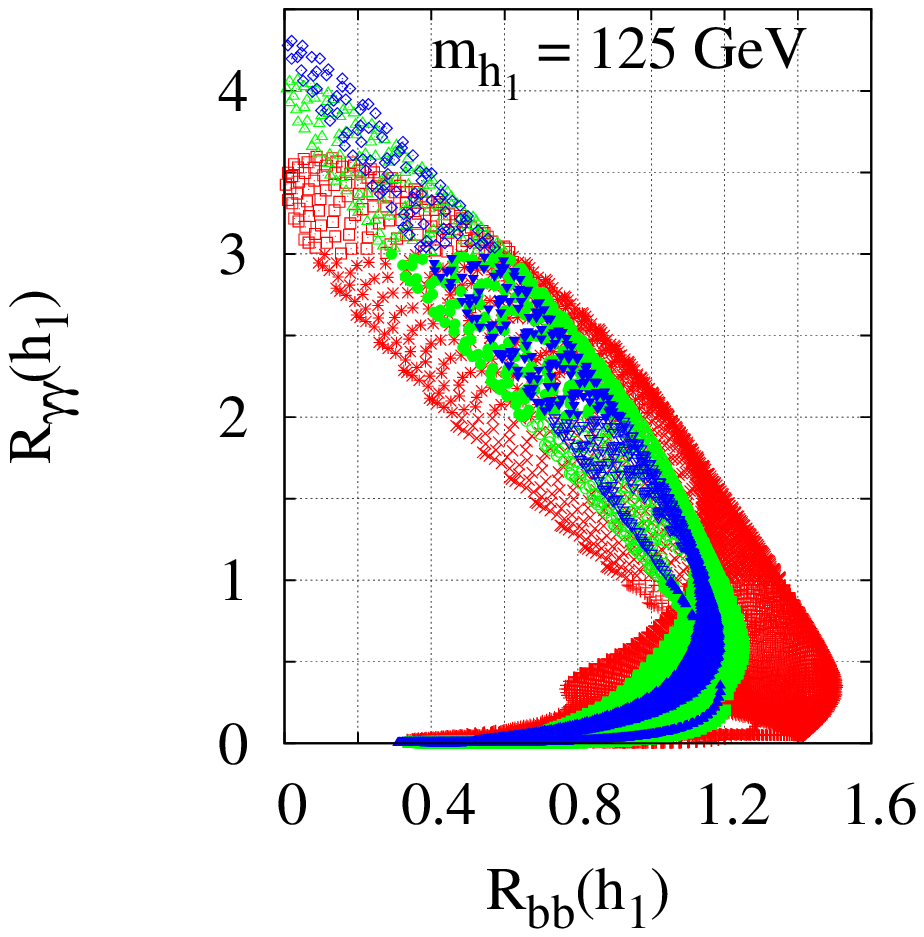}}
\end{tabular}
\caption{The correlation of $R_{WW}(h)$ and $R_{\gamma\gamma}(h)$ (left) and
  $R_{bb}(h)$ and $R_{\gamma\gamma}(h)$ in the c2HDM.}
\label{figu:5}
 \end{figure}
\section{Result}
In this section we present our numerical analysis for various Higgs decays.
We consider only the type II c2HDM.
In our analysis, we use the following 8 physical real parameters 
\cite{ElKaffas}:
\begin{equation}\hspace*{2.5cm}
\bigg\{M_{H_1},~~ M_{H_2},~~ \alpha_1,~~ \alpha_2,~~ \alpha_3,~~
M_{H^+},~~\tan \beta,~~ \Re e\, m^2_{12} \bigg\}.\label{set2}
\end{equation}

To satisfy $b\to s \gamma$ constraint in this model we assume that
 the charged Higgs mass is heavier than 295 
 GeV ~\cite{Ref:bsgNLO,Ref:bsgNLO2,Ref:bsgNNLO,Ref:bsgNNLO_THDM} (eventually
 the unitarity constraint excludes high values of $M_{H^\pm}$).  
In our scans, we assume $\tan\beta>1$ 
in order to satisfy $Z\to b\bar{b}$, 
B-B mixing constraints.
We also impose perturbative unitarity and vacuum stability 
constraints on the parameters of the scalar potential 
\cite{theor.constr.}  as well as the constraint coming from
electroweak physics, related to the precise determination of the
$\rho$-parameter~\cite{PDG}.

At collider one measures the total cross-section
$\sigma^{\gamma\gamma}_{h}=\sigma(pp\to h \to \gamma \gamma)$ where
the largest contribution to the production cross-section for this
observable $\sigma^{\gamma\gamma}_{h}$ is through gluon fusion, $gg \to h \to
\gamma \gamma$.  For phenomenological purpose, we will use the ratio
$R_{\gamma\gamma}$, $R_{WW}$ and $R_{b\bar{b}} $ defined
previously in equation~\ref{ratioxx}. 
Several recent studies use LHC data to put bounds on those ratios 
$R_{\gamma\gamma}$, $R_{WW}$ and $R_{b\bar{b}}$
in a model independent way \cite{Espinosa:2012ir}. We will not apply those 
bounds here but rather give the predictions of the c2HDM.
%

In Figure~\ref{figu:1}, we show for $m_{h_1}=125$ GeV 
 the  allowed parameter space in the $\alpha_1 - \alpha_2$ plane, for
  $\tan\beta = 1.5 $ (left), $\tan\beta = 2$ (middle) and $\tan\beta = 2.5$
  (right). The other parameters are taken as follow: $m_{h_1} = 125$ GeV,
  $m_{h_2} = 220$ GeV,  $m_{H^\pm} = 350$ GeV, $Re(m^2_{12}) = 200$ GeV,
  $\alpha_3 = \frac{\pi}{3}$.
The white region is excluded by one of the above constraints. 
It is clear that for small $\tan\beta=1.5$, the allowed region of 
$\alpha_1$-$\alpha_2$ plan is larger and get reduced when taking 
$\tan\beta = 2.5$. The coding color correspond to: 
$R_{\gamma\gamma} \le 1$  (red), $1\le R_{\gamma\gamma} \le 2$ (green), 
$2\le R_{\gamma\gamma} \le 3$ (blue) and $3\le R_{\gamma\gamma}$ (mangeta).
The dependence of the allowed regions on $R_{\gamma\gamma}(h_1)$ 
is illustrated in
figure~\ref{figu:1} while $\alpha_3$ kept fixed at $\pi/3$.
The figure shows the deviation of the c2HDM from the SM predictions can
occur due to the mixing effect appears at the tree-level.
For small $\tan\beta$ there are still regions where theoretical and
experimental constraints are satisfied. However, when $\tan\beta$ approaches
2.5 (or even 3), the allowed regions tend to be restricted to small values of
$\alpha_2$.  

In figure~\ref{figu:2}, we illustrate the ratio $R_{\gamma\gamma}$ as 
a function of $\alpha_1$ (left) and $\alpha_2$ (right). It is clear 
from this plot that enhancing $R_{\gamma\gamma}$ requires 
 $\alpha_1\approx \pm \pi/2$ and $\alpha_2\approx 0$. 
In fact taking $\alpha_1\approx \pm \pi/2$ and $\alpha_2\approx 0$ would 
suppress the partial width of the Higgs $\Gamma(h_1\to b\bar{b})$ which 
contribute dominantly to the total width. Such a suppression of the total
width will give an enhancement of $Br(h_1\to \gamma\gamma)$. This is clearly 
illustrated in figure~\ref{figu:3} where we plot $Br(h_1\to xx)$ as a function
of $\alpha_2$ for $\alpha_1 \approx \pi/2 $ rad, $\alpha_3 = 1.04$ rad, 
$\tan\beta = 1.5$ and $m_{h_1}= 125$ GeV. 
It is clear from this plot that when $\alpha_2$ decrease from $\alpha_2=-\pi$
down to  $\alpha_2\approx 0$ one can see from one side a reduction of $Br(h_1\to
b\bar{b},\tau^+\tau^-)$ and from the other side we can see
 an enhancement of $Br(h_1\to c\bar{c}, gg, W^+W^-, ZZ, \gamma\gamma)$.
 One can use CMS and ATLAS data to put some constraints on
  c2HDM parameter such as $\alpha_1$, $\alpha_2$ and $\tan\beta$. For 
$m_h=125$ GeV it is known from CMS data that $R_{\gamma\gamma}\leq 3.6$.
As one can see from figure~\ref{figu:2},
 this constraint on $R_{\gamma\gamma}$ clearly disfavor 
$\alpha_1\approx \pm\pi/2$, $\alpha_1\approx 0$ and $\tan\beta>2$.

As stated before, the Higgs in this model is a mixture of CP-even and CP-odd
 components. In Figure~\ref{figu:4} we plot the ratio $R_{\gamma\gamma}$ of
 the $h_1$ as a
 function of $|Z_{11}|^2=|(h_1WW)/h_{sm}WW|^2$  which is the 
coupling $h_1WW$ normalized to SM coupling. It is clear that
$R_{\gamma\gamma}(h_1)$ is enhanced only for $|Z_{11}|^2\geq
0.5$ which means that  $h_1$ has a substantial contribution 
from CP-even component.

  Since the main contribution to $h_1\to\gamma\gamma$ comes from the
  diagrams containing W-boson loops, a strong correlation between these two
  channel is expected. This correlation is confirmed by
  Figure~\ref{figu:5}(left). It is apparent from Figure~\ref{figu:5}(left) 
  that we have an approximate linear correlation between 
  $R_{\gamma\gamma}(h_1)$ and $R_{WW}(h_1)$. 
   The simultaneous enhancement of those two modes can be
 understood as an effect of large suppression of the main coupling 
 $h_1 b\bar{b}$ which favors both decays equally.

In this scenario, 
 enhancing/suppressing $R_{\gamma\gamma}(h_1)$ would also enhance/suppress
 $R_{WW}(h_1)$. 
 From Figure~\ref{figu:5}(right) one can see that 
large $R_{\gamma\gamma}(h_1)$  prefer rather suppressed $R_{bb}(h_1)$, but there is 
also substantial region of parameter space where we can have some excess 
in $R_{\gamma\gamma}(h_1)$ while $R_{bb}(h_1)$ is more or less consistent with SM.

\section{CP-Asymmetry of $h_1 \gamma \gamma$ mode}
Now we discuss the effect of the CP violating phases on some CP violation
observable related to di-photon events.
To quantify the size of CP violation in $h_1 \gamma \gamma$,
we introduce the following observable \cite{Choi:2004kq} :
\begin{eqnarray}
A_{+-} = 
\frac{\Gamma(h_1 \to \gamma\gamma(++))-\Gamma(h_1 \to \gamma\gamma(--))}
{\Gamma(h_1 \to \gamma\gamma(++))+\Gamma(h_1 \to \gamma\gamma(--))}
\end{eqnarray}
which is the rate difference between left-circular and right-circular
polarization of the final photons.

In order to get CP violation using the above 
asymmetries, we need both weak CP violating phases in 
the Lagrangian and CP conserving phases (strong phases) in
the absorptive parts of the one-loop amplitudes. The weak phase comes
from Higgs fermion pairs coupling that contribute to $h_1 \gamma \gamma$,
The CP conserving phases originate from various on-shell intermediate states
of the one-loop amplitudes. In this case, the strong
phases coming from cuts of $h_1\to b\bar{b}, \tau^+\tau^-$.

\begin{figure}[!ht]
\begin{tabular}{c}\hspace*{-1cm}
\resizebox{90mm}{!}{\includegraphics{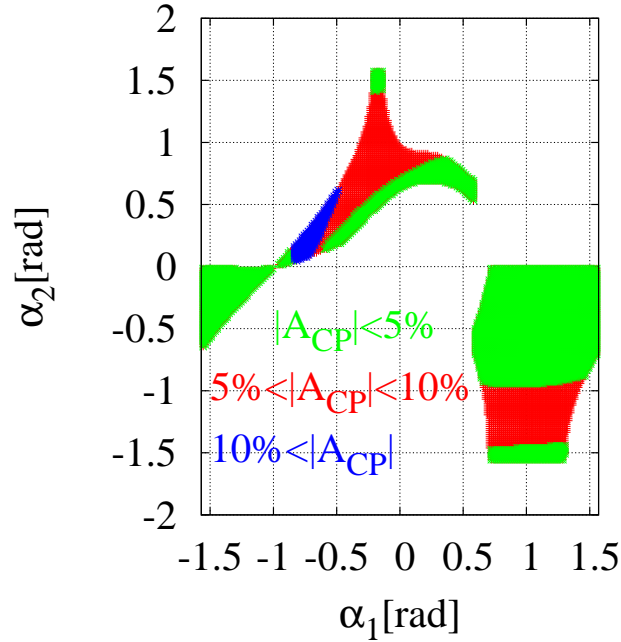}} 
\end{tabular}
\caption{The CP-Asymmetry $A_{+-}$ of $h_1\to \gamma\gamma$ mode in the ($\alpha_1 - \alpha_2$) plan with
  $m_{h_1} = 125$ GeV and $\tan\beta = 1.5$ in the  c2HDM. The other
  parameters are the same as in Figure.1}
\label{figu:6}
 \end{figure}

In Figure~\ref{figu:6}, we provide a scan for the rate difference between 
left-circular and right-circular polarization of the final photons
$A_{+-}$  as a function of $\alpha_1$ and $\alpha_2$. This asymmetry could
reach up to more than 10\% in some region of parameter space.
Such asymmetry can be measured in the photon-photon option of the 
International linear collider (ILC).

\section{Conclusions}

We have studied Higgs production and its decay to di-photon in c2HDM
in which the couplings of non-self-hermitian terms in the Higgs 
potential are complex. Due to the complex couplings, CP-even and 
CP-odd scalars 
will mix and the
mixture leads to CP violation. We find that a large region of 
the c2HDM parameter
space will be excluded when the constrains from precision measurement of
$\rho$ parameter and tree-level unitarity of Higgs-Higgs scattering are
included. With the allowed values of CP mixing angles, we found 
that the excess in $h_1\to \gamma \gamma$ process reported by ATLAS 
 and CMS at $M_{h_1} = 125$
GeV could be explained. We have also shown that such excess in $h_1\to \gamma
\gamma$ could be easily obtained with suppressed $h_1\to b\bar{b}$ but we
could also have an excess in $h_1\to \gamma
\gamma$ while $h_1\to b\bar{b}$ is still consistent with the SM 
predictions.\\ 
Additionally, with polarized photon we also 
study the direct CPA in di-photon channel. We find that the CPA could be as 
large as $10\%$, which could be tested at linear collider.\\

\noindent
{\bf{Note added:}} While we were finishing this work, we received papers 
\cite{pap01} and \cite{pap02}. 
\cite{pap01} deals with similar subject and our results agree in type II 
c2HDM which was considered in this study.
\cite{pap02} discuss production and detection of light charged Higgs 
 in type I 2HDM and also the effect of such light charged Higgs in $h\to
 \gamma\gamma$, we are not interested in such a scenario.


\centerline{\bf Acknowledgments}
\smallskip
A.A thanks the authors of \cite{pap01} for useful discussions.
The work of R.B was supported by the Spanish Consejo Superior de
Investigaciones Cientificas (CSIC).
This work is partially supported by the National Science Council of R.O.C. under Grant \#s: NSC-100-2112-M-006-014-MY3 (CHC).

\end{document}